\documentstyle[12pt,aaspp4] {article}

\begin{document}

\title{ROSAT Timing of the LMC Pulsar 0540-69\footnote{This research has made use of data obtained from the High Energy Astrophysics Science Archive Center (HEASARC) provided by NASA's Goddard Space Flight Center}}

\author{Stephen S. Eikenberry\altaffilmark{1}, Giovanni G. Fazio, Scott M. Ransom}
\affil{Harvard-Smithsonian Center for Astrophysics, Cambridge, MA 02138}

\altaffiltext{1}{Current address: Downs Laboratory, 320-47, California Institute of Technology, Pasadena, CA  91125}

\begin{abstract}

	We present a timing study of the young rotation-powered pulsar
0540-69 in the Large Magellanic Cloud, based on 130 kiloseconds of
archival ROSAT data spanning a $\sim 3$-year period.  We use ``$f-
\dot f$'' techniques to measure the pulsar frequency as a function of
frequency derivative at 17 independent epochs.  From these
measurements we derive a timing solution with a braking index $n =
2.5^{+0.6}_{-0.7}$, and we compare this solution to previous timing
studies of 0540-69.  Using this frequency-based solution, we create 27
pulse profiles and perform a time-of-arrival (TOA) analysis to
investigate further the pulsar's timing behavior.  While we can
successfully fit smooth spin-down models to subsets of the TOAs
spanning up to 2 years, we are unable to obtain acceptable
phase-coherent fits to the entire 3-year set of TOAs.  This behavior
provides the first clear evidence for timing noise in 0540-69.  We
discuss the implications of these results for understanding previous
studies of the timing behavior of 0540-69.

\end{abstract} 


\section{Introduction}

	The pulsar 0540-69 in the Large Magellanic Cloud (LMC) is one
of the youngest and most luminous rotation-powered pulsars, with a
spin-down age of $t_{sd} \sim 1500$ yrs and a spin-down luminosity of
$L_{sd} \sim 10^{38} {\rm erg}/{\rm s}$.  The pulsar was first
discovered as a 50-ms X-ray pulsar by Seward, Harnden and Helfand
(1987) using the Einstein IPC, with follow-up ground-based
observations revealing optical pulsations of magnitude $m_V \sim 22.5$
(Middleditch and Pennypacker, 1985).  Despite the large spindown
luminosity of 0540-69, it is a faint radio pulsar, with a 640 MHz flux
of only $\sim 0.4$ mJy (Manchester et al., 1992), requiring long
integrations with a 64-meter telescope for detection.  It is these
observational properties of the pulsations which have historically
made the timing of 0540-69 a difficult task: in the radio and optical
regimes it requires large time allocations on large telescopes for
extended periods.  While X-ray observations can detect 0540-69
readily, the sheer scarcity of satellite time limits dedicated study
in this band.  Thus, many of the timing observations of 0540-69 have
been conducted as ``add-ons'' to larger programs (such as X-ray and
optical searches for a pulsar in SN1987A).

	In addition to timing observations of 0540-69 being relatively
scarce, the published timing studies have produced differing, often
contradictory results.  One of the most interesting measurements is of
the pulsar braking index
\begin{equation} 
n = {f \ddot f \over{(\dot f)^2}} 
\end{equation} 
where $f$ is the pulsar frequency and $\dot f$ and $\ddot f$ are its
first and second derivatives with respect to time.  The braking index
is a key indicator of the pulsar's magnetic field geometry, and
magnetospheric processes.  The first reported measurement for 0540-69
was $n=3.6 \pm 0.8$ by Middleditch, Pennypacker, and Burns (1987).
Since then various groups have reported values of $n$ ranging from 2.0
to 2.7, usually with small error bars excluding the results of other
groups (see Table 1).  Even when the braking indices agree, the
frequency (and derivative) values differ so significantly that such
agreement is likely a coincidence (see e.g. Nagase et al., 1990 and
Manchester and Peterson, 1989).  Furthermore, in order to explain
these discrepancies, virtually every paper listed in Table 1 has
claimed that the results prior to its own were contaminated by
glitches, or at least that a glitch occurred between the observations.
Thus the timing behavior of 0540-69 is hardly a settled matter, more
than a decade after the pulsar's discovery.

	In this paper, we present a new timing analysis of 0540-69
utilizing a unique and previously untapped resource for this work -
the ROSAT data archive.  Due to the wide field of the ROSAT PSPC
($\sim 1^{\rm o}$ radius), 0540-69 is in the field-of-view for a large
fraction of the pointings towards the LMC (see Figure 1).  Because
there was a large number of programs studying a wide range of objects
in the LMC (i.e. LMC X-1, SN1987A, supernova remnants, etc.), the
total time during which 0540-69 was in the PSPC field comes to nearly
160,000 seconds.  Previously, most of these data have been difficult
to use for timing of 0540-69 due to the nature of scheduling X-ray
observations -- typically the pulsar was observed for several hundred
seconds at a time, with gaps of hours, days, or even weeks until the
next pointing.  Over such long timespans the effects of the pulsar's
$\dot f$ become very important, and traditional methods (e.g. FFTs)
are inadequate for performing the timing analysis over these gaps.
However, the individual small exposures alone often do not produce
sufficient signal-to-noise for timing analysis.  Thus, for this work
we employ more advanced ``$f- \dot f$'' techniques (Ransom and
Eikenberry, 1997; Eikenberry, 1997) to determine the Fourier power as
a function not only of frequency $f$, but also of the frequency
derivative $\dot f$ -- that is, $P(f, \dot f)$.  Such techniques are
suitable for pulsar timing with ``gappy'' data, and thus allow us to
analyze the unique ROSAT PSPC archival data set for timing of 0540-69.

	In the following sections, we first discuss the archival X-ray
observations from ROSAT and the data reduction techniques used to
determine the pulsar's Fourier power spectrum for each observation,
$P(f, \dot f)$.  Next, we present the pulsar timing solution as
determined by the analysis of the various $P(f, \dot f)$ measurements,
and we compare this solution to several previous determinations.  We
then move on to a refined timing solution using pulse time-of-arrival
(TOA) analyses.  Next, we discuss the implications of these results
for understanding the timing behavior of 0540-69 in the context of
prior work.  Finally, we present our conclusions.

\section{Observations and Data Reduction}

	We selected the archival data for this work using the
WWW-based archival search software of the HEASARC project at NASA
Goddard Space Flight Center.  We chose data sequences for which
0540-69 was within a $30 \arcmin$ radius of the field center, having
determined empirically that the deterioration of the PSPC point spread
function (PSF) for radii greater than $30 \arcmin$ resulted in
unacceptably low signal-to-noise ratios.  We also selected only those
sequences which had total on-source exposure times of $t_{obs} > 2500$
s, as shorter total exposures usually resulted in individual exposures
too short to be useful.  We list the selected data sequences in Table
2.

	We used the PROS software package within IRAF to create images
of the field (see Figure 1 for an example) and select all the photon
events within a circular aperture centered on 0540-69, varying the
aperture radius depending on the pulsar PSF in each observation.  We
then created time-sorted event lists and light curves for each
sequence.  For easier processing, we subdivided the sequences into
"pieces" with maximum durations of $t_{obs} < 2$ weeks (see Table 3
for a listing).  We then corrected the photon event times to the solar
system barycenter using a pulsar position of ${\rm RA = 05^h \ 40^m \
11^s.03}$ and ${\rm Dec} = -69^{\rm o} \ 19\arcmin \ 57\arcsec .5$
(J2000)\footnote{Deeter, Nagase, and Boynton (1997) use a different
position for the pulsar, offset by $\sim 4$ arcsec.  While this offset
is significant for the TOA analysis (as we discuss in the next
section), it does not significantly affect the frequency results here
and in section ~\ref{freqsol}.}  (Gouiffes, Finley, and \"{O}gelman,
1992 -- hereafter GF\"{O}), and we exported these photon arrival
times outside IRAF for the pulsation analysis.

	We calculated the pulsar Fourier power spectrum for each piece
in Table 3 using the signal-folding technique for ``$f - \dot f$''
analysis (Ransom and Eikenberry, 1997; Eikenberry, 1997; Eikenberry et
al., 1997).  Since this technique has been described elsewhere, we
only briefly summarize it here.  The effect of a frequency derivative
or a frequency offset between the Fourier frequency and the pulsar
frequency is negligible over a small time interval, but it can have a
significant impact on the resulting Fourier power over an entire time
series.  Thus, we subdivide the pulsar time series into segments of
length $\sim 4$ seconds and calculate the Fourier transform over each
segment, $F_i$.  The FFT algorithm gives a Fourier amplitude equal to
the complex sum of the $F_i$.  A frequency derivative or frequency
offset will cause each $F_i$ to suffer a phase shift relative to its
neighbors, so that when added they are slightly out of phase -- the
$F_i$ vectors are rotated in the complex plane -- and power is lost.
However, this effect can be countered by ``derotating'' the $F_i$
through multiplication by a complex phase factor corresponding to the
effect of the frequency derivative and offset, so that the resulting
$F_i^{(rot)}$ will add in phase and power will be recovered.
Therefore, for a given frequency offset $w=f-f_0$ and a given
frequency derivative $\dot f$, we multiply the $F_i$ by the phase
factors
\begin{equation}
\nonumber
e^{-i \phi_i} , \ \ \phi_i = w t_i + {1 \over{2}} \dot f {t_i}^2 ,
\end{equation}
where $t_i$ is the time elapsed from the beginning of the observation
to the $i^{th}$ segment.

	We selected an initial frequency for signal folding by taking
a small segment of data, performing an FFT on it, and taking the
frequency from the largest Fourier peak near $f \sim 19.85$ Hz.  We
then ``folded'' the time series at this initial guess frequency,
saving the complex Fourier amplitudes for every 4.096-second segment
of the time series.  Next, we ``derotated'' this Fourier series for a
range of frequency and frequency derivative combinations, and
calulated the Fourier power at each combination -- $P(f, \dot f)$.  We
varied the range and step size for $f$ according to the duration of
the data "piece", but we used a uniform set of $\dot f$ values from
$-1.9 \times 10^{-10}$ Hz/s to $-1.85
\times 10^{-10}$ Hz/s with a step size of $\Delta \dot f = 1 \times
10^{-14}$ Hz/s.  We repeated these steps for the next 4 higher
harmonics of the fundamental frequency, and added the resulting powers
to obtain the sum power, $P_{sum} (f, \dot f)$.  Finally, using
$\sigma_P = \sqrt{2 P}$, for the $j^{\rm th}$ "piece" at each value of
$\dot f$, we calculated the best-fit initial frequency $f_{0,j}(\dot
f)$ and the frequency uncertainty $\sigma_f(\dot f) $\footnote{For
signals with window functions sufficiently close to a simple square window,
the frequency can be determined with an uncertainty smaller than this
by a factor of $\sim 2-3$ for 0540-69 (see Ransom and Eikenberry,
1997; Eikenberry, 1997).  However, for several ``pieces'' here, the
window function is sufficiently different that this is not true.  Thus
for the sake of consistency we use this estimate for $\sigma_f$,
knowing that it is an over-estimate in several cases.  Note also that
the effects of the frequency second derivative $\ddot f$ on
$f_{0,j}(\dot f)$ are negligible over the maximum duration of each
individual "piece".}.

\section{Timing Analysis}

\subsection{Frequency-based Timing Solution} \label{freqsol}

	Once we had the starting time for the data ``pieces'' and their
frequency estimates $f_{0,j}(\dot f)$, we calculated a timing solution
for the pulsar of the form
\begin{equation}
f(t) = f_0 + \ \dot f t \ + {1 \over{2}} \ddot f t^2 \ \ .
\end{equation}
To do so, we looped over a range of $\ddot f$ and the $\dot f$ values
above, calculating the least-squares value of $f_0$ (the pulsar
frequency at the beginning of the first data "piece" -- MJD
48059.7864091) for the data, and the corresponding value of $\chi^2$.
We found a minimum $\chi^2$ value of 1.9 for 15 degrees of freedom (or
$\chi^2_\nu = 0.13$)\footnote{Note that the procedure above may
overestimate $\sigma_f$ by a factor of up to 2-3, resulting in a
correspondingly low value of $\chi^2_{nu}$.} for the parameters
listed in Table 4.

	We compare this timing solution to those of previous works in
Figure 2.  We find excellent agreement between our solution and the
ephemeris of Deeter, Nagase, and Boynton (1997) (hereafter DNB).  We
see a significant discrepancy with the ephemeris of GF\"{O} -- the
difference between their ephemeris and our local frequency
measurements results in $\chi^2 = 1422$ for 17 degrees of freedom.  We
find very large discrepancies between our solution and the
extrapolation of the timing ephemerides of \"{O}gelman and Hasinger
(1990) and Manchester and Peterson (1989).

\subsection{TOA-based Timing Solutions}

\subsubsection{TOA determination}

	We next moved to a time-of-arrival (TOA) timing analysis for
pulsar 0540-69 in order to refine the timing solution above.  TOA
analysis links the individual observations in phase, providing a much
more sensitive probe of the pulsar's timing behavior.  In order to
carry out this analysis, we first created pulse profiles.  For added
timing sensitivity, we subdivided some of the data ``pieces'', resulting
in a total of 27 pulse profiles\footnote{We excluded the HRI data
sequence RH150008, as the different energy response of the HRI may
produce systematic differences in the observed pulse shape, and thus
systematic phase offsets for this analysis.}.  We folded the pulse
profiles using the timing solution in Table 4, with 1 ms phase bins.

	We next created a master pulse profile as follows.  First, we
correlated the first pulse profile against the second, shifting the
second profile bin-by-bin to obtain the correlation coefficient as a
function of phase shift.  We then added the shifted second profile to
the first profile at the phase shift which produced the largest
correlation coefficient.  We then correlated the sum profile against
the third profile, adding the shifted third profile to the sum as we
did for the second profile, and we repeated this procedure for the
$4^{\rm th}$ through $27^{\rm th}$ profiles.  This resulted in a
temporary master profile.  However, in creating this temporary master,
each profile was correlated against a different sum profile than all
the other profiles, leaving room for systematic errors.  In order to
remove these potential errors, we created the final master pulse
profile by correlating all of the individual profiles against the
temporary master, shifting each bin-by-bin, and adding all the shifted
profiles using the shifts which produce the largest correlation
coefficient with the temporary master profile.  We show the resulting
master pulse profile in Figure 3.  We also compared the individual
pulse profiles to the master profile in order to search for long-term
chnages in the pulse shape.  We found no significant evidence of such
variability.

	We next determined the individual pulse TOAs as follows.  We
determined the pulse phases by correlating the profiles against the
master pulse profile in Figure 3, taking the phase as the bin-shift
which produced the largest correlation coefficient.  We determined the
uncertainties in the phases (and thus in the TOAs) by a Monte Carlo
simulation.  First, we assumed that the errors in the pulse profiles
were dominated by the Poissonian uncertainties in the number of events
in each phase bin.  For a given pulse profile, we added to each phase
bin a random number drawn from a distribution corresponding to the
uncertainty in the number of events in the phase bin, resulting in a
simulated pulse profile.  We then correlated this simulated profile
against the master profile, giving the simulated pulse phase.  We then
repeated this procedure 1000 times, taking the uncertainty in the
pulse phase to be the standard deviation in the simulated pulse
profiles plus the systematic uncertainty of $\pm 0.5$ phase bins.  The
pulse TOA is then the start time of the observation plus the pulse
phase multiplied by the pulsar period.

\subsubsection{Timing solutions -- I}

	In order to determine the timing solution given the pulse
TOAs, we simply calculated the $\chi^2$ for the phase residuals to a
solution of the form
\begin{equation} 
\label{mod1}
\phi (t) = \phi_0 + f_0 t + {1 \over{2}} \dot f_0 t^2 + {1 \over{6}} \ddot f t^3 
\end{equation}
We resorted to an iterative process in order to calculate the $\chi^2$
values over the entire range of parameter space.  First, we selected a
segment of our timespan which is particularly dense with TOA
determinations -- covering TOAs 13-23.  We then looped over the range
of allowed solutions from Table 4 using the step sizes $(df_0, d \dot
f_0, d \ddot f)$ dictated by the $\sim 5 \times 10^6$-second timespan
of this subset of our observations.  We then expanded the subset to
include TOAs 10-23, looping over the allowed solutions from the 13-23
subset with smaller step sizes corresponding to the timespan of the
10-23 data set.  We repeated this procedure for the TOA 1-23 subset,
and finally the TOA 1-27 subset.

	For the subset of TOAs 1-23, we found 2 different timing
solutions, A and B in Table 5, both with chi-squared values of $\chi^2
\simeq 19$ for 19 degrees of freedom.  The difference in these
solutions is due to ambiguities in the pulse cycle count over some of
the large gaps in the ROSAT coverage of 0540-69.  Note that the
$\chi^2$ values for both solutions are very good.  However, when we
extended the analysis to the full set of TOAs, we found $\chi^2_{min}
= 39.3$ for 23 degrees of freedom -- a significantly poorer fit (see
Figure 4).

	This poorer fit initially suggested the presence of a glitch
between TOA 23 and TOA 24, so we repeated the determination of a
timing solution starting from TOAs 10-23 and extending the subset to
TOAs 10-27.  We found a family of 4 solutions with $\chi^2_{min} =
16.8$ for 14 degrees of freedom (solutions C-F in Table 5), showing
that it is possible to fit a model of the form of eqn. ~\ref{mod1}
over this timespan\footnote{Again, the differences in solutions C-F in
Table 5 are due to pulse numbering ambiguities over the larger gaps in
the ROSAT coverage of 0540-69}.  While this does not preclude the
presence of a glitch between TOA 23 and TOA 24, it does make it
impossible to determine conclusively that a glitch did occur.
However, when we extend solutions C-F to the full set of TOAs, we
again find a poor fit ($\chi^2_{min} = 39.4$ for 23 dof).

\subsubsection{Timing solutions -- II}

	The above analysis assumed a position for 0540-69 determined
by X-ray imaging observations.  However, DNB have recently suggested
that a more accurate position for the pulsar is RA = $5^h
\ 40^m \ 11^s.57$, Dec = $-69^{\rm o} \ 19 \arcmin \ 54 \arcsec .9$ (J2000), as
determined by timing observations.  This $\sim 4\arcsec$ shift in
position will cause a systematic shift in the observed barycentric
TOAs consisting of a sine wave with a 1 year period and an
amplitude of $\sim 10$ ms -- significant variations given our typical
TOA uncertainties of $\sim 2-4$ ms.  (Note that such a sine wave would
produce systematic frequency offsets $\delta f < 2 \times 10^{-9}$ Hz
-- much less than the statistical uncertainties in the individual
frequencies used for the solution in Table 4).  In order to check if
these effects were responsible for the poor fits seen above, we
calculated a revised set of TOAs using the position of DNB.

	We then repeated the timing analysis as above for the revised
TOAs.  For the subset of TOAs 1-23, we found a family of 3 solutions
with $\chi^2_{min} = 18.6$ for 19 dof (solutions G-I in Table 5 ) -
again an excellent fit.  However, when we extended these solutions to
the full set of revised TOAs, we again found a much poorer fit -
$\chi^2_{min} = 44.5$ for 23 dof.  For the subset of TOAs 10-27, we
also found 5 acceptable solutions (J-N in Table 5) with $\chi^2_{min}
= 14.3$ for 14 dof.  When we extended these solutions to TOAs 1-27 we
found a solution with $\chi^2 = 32.1$ for 23 degrees of freedom (see
solution O in Table 5).  While this solution is better than those we
found previously, we note that the chi-squared increased by $\Delta
\chi^2 = 17.8$ for TOAs 1-9.  Thus, while TOAs 10-27 may follow this
solution, we can reject the hypothesis that the first 9 TOAs follow
the same solution at the 95\% confidence level.

\section{Discussion} \label{discuss}

\subsection{Frequency-determined Solutions}

	As we saw in Figure 2, our frequency-determined solution in
Table 4 provides an acceptable match with the ephemeris of DNB, and
also with that of Nagase et al. (1990).  DNB is a continuation of the
Nagase et al. (1990) work by several of the same authors, and expands
the GINGA-based X-ray timing of Nagase et al. (1990) to include more
GINGA data along with optical data and a small amount of
previously-published ROSAT data (Finley et al., 1993).  The fact that
our independent timing solutions agree to such high accuracy implies
that they do indeed represent the frequency behavior of 0540-69.

	We find a significant discrepancy between the solution in
Table 4 and the timing ephemeris of GF\"{O}.  GF\"{O} noticed a
similar mismatch with the ephemeris of Nagase et al. (1990), and
hypothesized that this discrepancy might be due to a third derivative
of the frequency with time, combined with the fact that the Nagase et
al. (1990) observations had an earlier average epoch than GF\"{O}.
However, the fact that we match the GINGA ephemeris at an even later
epoch seems to contradict this assessment.  When we compare our
frequency-determined solution to the local frequency measurements of
GF\"{O}, we find excellent agreement (though the typical uncertainties
in the GF\"{O} measurements are larger than the discrepancy seen in
Figure 2). Thus, the discrepancy seems to arise in the TOA-based
solution for the GF\"{O} data.

	In comparing our frequency-based timing solution to the
ephemerides of Manchester and Peterson (1989) and \"{O}gelman and
Hasinger (1990), we find very large discrepancies -- similar to those
found by Nagase et al. (1990), GF\"{O}, and DNB.  This adds further
weight to the argument that the Manchester and Peterson (1989)
ephemeris is either erroneous or affected by timing noise -- in a
reanalysis of the Manchester and Peterson (1989) TOA data, DNB claim
that there is evidence for a large glitch in the range MJD
46900-47000.

	It is interesting to note that our frequency-based solution in
Table 4 agrees with the local frequency measurements of both GF\"{O}
and \"{O}gelman and Hasinger (1990).  The discrepancies all occur
between our frequency-based solution and the long-term TOA-based
solutions\footnote{DNB use phase-coherent TOA analysis only over short
timespans, due to clock problems with GINGA.  See their paper for
details.}.  This fact may be explained if significant timing noise is
present in the timing behavior of 0540-69.  By way of comparison, for
several pulsars with timing noise (i.e. 0832+26 and 0736-40, as
discussed in Cordes and Downs, 1985), the frequency residuals $\delta
f$ due to timing noise scale roughly as $\delta f \sim {\delta \phi
\over{T}}$, where $T$ is the timespan of monitoring and $\delta \phi$
are the phase residuals due to timing noise.  Thus, over a timescale
of $\sim 1-2$ years, similar timing noise in 0540-69 could produce
phase residuals of several cycles -- clearly large enough to alter
TOA-based solutions -- while producing frequency modifications of only
$\delta f \sim 10^{-7}$ Hz -- too small to affect any of the
frequency-based solutions we discuss here.

\subsection{TOA-determined Solutions}

	We have seen above that we are able to fit successfully large
subsections of the ROSAT TOA data set with smooth timing models of the
form of eqn. ~\ref{mod1}.  However, regardless of which pulsar
position we assume and which subset of the TOAs we use to determine a
preliminary solution, we are unable to produce a phase-coherent fit to
the entire data set.  This behavior strongly suggests the presence of
timing noise which is contaminating the fits determined from the pulse
TOAs.  The presence of such noise in the timing behavior of pulsar
0540-69 may also explain the incompatibility of previous timing
studies with one another, as some or all of them may be contaminated
by timing noise.  While previous workers have suggested such a
situation to explain their poor matches with other timing solutions,
the incompatibilities have always been {\it between} different
studies.  Thus, there has always remained speculation that the
observed incompatibilities are due to differences in analysis and
handling of the data from one study to the next.  However, this work
has presented the first evidence of timing noise within a single
self-contained study, and thus eliminates differences in the treatment
of the data as a possible explanation for the timing discrepancies.

	Solution O in Table 5, while only marginally unacceptable,
gives a value for the braking index of $n \sim 2.5$ -- significantly
different from previous determinations.  Thus, even if we were to
``stretch'' our definition of a ``good fit'', we would be left to
conclude that the estimated braking index for 0540-69 varies with
time.  As Cordes and Downs (1985) note, such time variability is a
prime indicator of the presence of timing noise.

	Given that timing noise is present in the timing history of
0540-69, we would like to determine the form that such noise takes.
Given a densely-sampled timing data set, large discontinuities
(glitches) would appear as ``jumps'' in the pulsar frequency and
frequency derivative, with exponential recoveries.  Even with somewhat
sparser coverage, glitches typically leave long-term frequency
derivative offsets of ${\Delta \dot f \over{\dot f}} \sim 10^{-4}$
(Lyne, Pritchard, \& Smith, 1993).  However, the sparseness of our
data set is so extreme, especially near TOAs 24-27, that we cannot
determine $\dot f$ to sufficient accuracy to look for such an offset
between the beginning and end of the data set.

	Another form of timing noise would be smaller discontinuities
-- the so-called ``random walk'' timing noise (Cordes and Downs, 1985).
However, determining the characteristics of such timing noise can be
very difficult even with a densely- and evenly-sampled timing data
set.  Given the uneven and sparse coverage of 0540-69 by ROSAT, we are
unable to distinguish between large and small discontinuities, and
thus we cannot hope to characterize the properties of any small
discontinuities if they are present.

\section{Conclusions}

	Using novel Fourier techniques, we have analyzed archival
ROSAT observations of the Large Magellanic Cloud for a $\sim 3$-year
timespan in order to determine the timing characteristics of the young
isolated pulsar 0540-69.  We present the conclusions from this
analysis below:

	1) Using local frequency measurements, we determine a
well-fitting timing solution with a braking index of $n = 2.5
^{+0.6}_{-0.7}$.  This timing solution is consistent with that of DNB,
but is inconsistent with the TOA-based timing results of GF\"{O},
\"{O}gelman and Hasinger (1990), and Manchester and Peterson (1989).

	2) Our frequency-based solution is consistent with the local
frequency measurements of GF\"{O} and \"{O}gelman and Hasinger
(1990).  This agreement may be understood in light of the discrepancy
with their TOA-based solutions if timing noise similar to that seen in
several other pulsars (e.g. Cordes and Downs, 1985) is present in the
timing behavior of 0540-69.

	3) Using the frequency-based timing solution for 0540-69, we
have constructed pulse profiles for the individual data segments and a
master pulse profile.  We find no evidence for time variability in the
pulse shape over the 3-year timespan of the ROSAT observations.

	4) Using the pulse profiles, we have determined pulse arrival
times (TOAs) for 0540-69, and created TOA-based timing solutions.
While we can successfully fit smooth spin-down models to subsets of
the TOAs spanning up to $\sim 2$ years, we are unable to obtain
acceptable phase-coherent fits to the entire 3-year set of TOA's.
This behavior indicates the presence of timing noise in 0540-69, and
is the first evidence for such timing noise within a single,
independent timing study.

	5) The best-fitting TOA-based timing solution, using the
pulsar position determined by DNB, is excluded at the 95\% confidence
level.  Even accepting this solution, comparing its braking index ($n
\sim 2.5$) to previous measurements would still require the presence
of timing noise in 0540-69.

	6) As a result of this timing noise into account, simple
estimates of the braking index for 0540-69 using TOA analyses which
ignore the timing noise are unreliable.

	7) Given the sparseness and uneven sampling provided by the
archival ROSAT coverage of 0540-69, we are unable to characterize the
nature of this timing noise.  Such work will require regular, frequent
optical/X-ray observations of 0540-69 over a timespan of several
years.


\begin{deluxetable} {lcc}
\tablecolumns{3}
\tablewidth{0pc}
\tablecaption{Previously-determined Braking Indices}
\tablehead{
\colhead{}		& \colhead{Time Span}	& \colhead{Braking Index}	\\ \colhead{Authors} & \colhead{(MJD)} & \colhead{(n)}}
\startdata
Middleditch, Pennypacker, \& Burns (1987) & 44186-46112 & $3.6 \pm 0.8$ \nl
Manchester \& Peterson (1989) & 46600-47400 & $2.01 \pm 0.02$ \nl
\"{O}gelman and Hasinger (1990) & 45924-46246 & $2.74 \pm 0.10$ \nl
Nagase et al. (1990) & 46900-47500 & $2.02 \pm 0.01$ \nl
Gouiffes, Finley, \& \"{O}gelman (1992) & 47540-48360 & $2.04 \pm 0.02$ \nl
Deeter, Nagase, \& Boynton (1997) & 46110-47725 & $2.080 \pm 0.003$ \nl
\enddata
\end{deluxetable}


\begin{deluxetable} {lccr}
\tablewidth{0pc}
\tablecaption{List of ROSAT Data Sequences}
\tablehead{
\colhead{}		& \multicolumn{2}{c}{MJD}	&
\colhead{Total Exposure} \\
\cline{2-3} \\
\colhead{Sequence}	& \colhead{Start}	& \colhead{End}	&
\colhead{(s)}}
\startdata
WP110167 & 48059 & 48059 & 2966 \nl
WP150044 & 48090 & 48136 & 5333 \nl
RH150008\tablenotemark{a}& 48097 & 48098 & 18126 \nl
WP400052 & 48307 & 48310 & 8953 \nl
WP600100 & 48367 & 48726 & 24989 \nl
RP500100 & 48535 & 48756 & 27207 \nl
RP400079 & 48589 & 49135 & 7666 \nl
RP500131 & 48741 & 48760 & 16069 \nl
RP500140A1 & 49084 & 49087 & 9993 \nl
RP500140A2 & 49158 & 49173 & 10766 \nl
\tablenotetext{a}{This sequence used the ROSAT HRI instrument.}

\enddata
\end{deluxetable}


\begin{deluxetable} {lcc}
\tablewidth{0pc}
\tablecaption{List of ROSAT Data ``Pieces''}
\tablehead{
\colhead{}		& \multicolumn{2}{c}{MJD}	\\
\cline{2-3} \\
\colhead{Piece ID}	& \colhead{Start}	& \colhead{End}}
\startdata
WP110167 & 48059 & 48059 \nl WP150044 a-f & 48090 & 48091 \nl RH150008
& 48097 & 48098 \nl WP400052 & 48307 & 48310 \nl WP600100 a-b & 48367
& 48371 \nl RP500100 a & 48535 & 48535 \nl RP400079 b-d & 48635 &
48635 \nl WP600100 f-j & 48703 & 48710 \nl WP600100 k-n & 48712 &
48718 \nl WP600100 o-p & 48723 & 48726 \nl RP500131 a-c & 48741 &
48741 \nl RP500100 b & 48742 & 48742 \nl RP500100 c-f & 48752 & 48756
\nl RP500131 d-s & 48753 & 48760 \nl RP500140A1 & 49084 & 49087 \nl
RP500140A2 a-f & 49158 & 49165 \nl RP500140A2 g-j & 49171 & 49173 \nl

\enddata
\end{deluxetable}


\begin{deluxetable} {lr}
\tablewidth{0pc}
\tablecaption{Frequency-based Timing Solution}
\tablehead{
\colhead{Parameter}		& \colhead{Value\tablenotemark{a}}}
\startdata
Epoch (MJD) & 48059.78640911 \nl
$f_0$ (Hz) & 19.8534877(18) \nl
$\dot f_0$ (Hz/s) & $-1.8886(8) \times 10^{-10}$ \nl
$\ddot f$ (${\rm Hz/s^2}$) & $4.5^{+1.2}_{-1.4} \times 10^{-21}$ \nl
Braking Index ($n$) & $2.5^{+0.6}_{-0.7}$ \nl

\tablenotetext{a}{Values in parentheses are the uncertainties in the last quoted digits.}

\enddata
\end{deluxetable}


\begin{deluxetable} {lrrrrrr}
\tablewidth{0pc}
\tablecaption{TOA-based Timing Solutions}
\tablehead{
\colhead{}	&\multicolumn{4}{c}{Parameters\tablenotemark{a}} &
\colhead{} & \colhead{} \\
\cline{2-5} \\
\colhead{} & \colhead{$f_0-f_{ref}$} & \colhead{$\dot f_0$} &
\colhead{$\ddot f$} & \colhead{} & \colhead{} & \colhead{} \\
\colhead{Sol'n}		& \colhead{($10^{-9}$Hz)}	&
\colhead{($10^{-10}$ Hz/s) } & 
\colhead{($10^{-21} {\rm Hz/s^2}$)}	& 
\colhead{Braking Index} & \colhead{TOA Range} & 
\colhead{Pos.}\tablenotemark{c}}
\startdata
A & 6456(10) & -1.888147(9) & 3.67(3) & 2.05(2) & 1-23 & 1 \nl
B & 6972(10) & -1.888279(8) & 3.88(3) & 2.16(2) & 1-23 & 1 \nl
C & 5580(15) & -1.88771(5) & 2.930(5) & 1.632(3) & 10-27 & 1 \nl
D & 6750(20) & -1.888231(7) & 3.855(11) & 2.147(6) & 10-27 & 1 \nl
E & 8872(12) & -1.8891(2) & 4.998(7) & 2.781(4) & 10-27 & 1 \nl
F & 9260(50) & -1.889071(17) & 5.24(3) & 2.92(2) & 10-27 & 1 \nl
G & 6547(10) & -1.888187(9) & 3.84(3) & 2.14(2) & 1-23 & 2 \nl
H & 6644(5) & -1.888235(5) & 3.955(16) & 2.202(9) & 1-23 & 2 \nl
I & 8900(1100) & -1.88886(19) & 4.762(11) & 2.650(6) & 1-23 & 2 \nl
J & 5280(30) & -1.887672(7) & 2.809(16) & 1.565(9) & 10-27 & 2 \nl
K & 6100(50) & -1.888016(17) & 3.52(3) & 1.96(2) & 10-27 & 2 \nl
L & 6490(40) & -1.888156(15) & 3.76(2) & 2.094(11) & 10-27 & 2 \nl
M & 8400(1100) & -1.88869(20) & 4.470(17) & 2.488(9) & 10-27 & 2 \nl
N & 10192(12) & -1.889163(5) & 5.041(8) & 2.804(4) & 10-27 & 2 \nl
O & 7286(5) & -1.888489(3) & 4.450(5) & 2.477(3) & 1-27 & 2 \nl

\tablenotetext{a}{Values in parentheses are the uncertainties in the last quoted digits.  Pos=1 solutions are referenced to MJD 48059.78640911.  Pos=2 solutions are referenced to MJD 48059.78640904.}
\tablenotetext{b}{$f_{ref} = 19.85348$ Hz}
\tablenotetext{c}{Pulsar position used: 1 - GF\"{O}, 2 - DNB}
\enddata
\end{deluxetable}

\acknowledgements

	We would like to thank the SAO PROS support group for help
with the ROSAT data reduction; F. Seward, J. Cordes, R. Narayan, and
J. Grindlay for helpful comments on an early draft of this work.
S. Eikenberry was supported by a NASA Graduate Student Researcher
Program fellowship through NASA Ames Research Center.

\vfill \eject

\begin{figure}
\vspace*{130mm}
\includegraphics{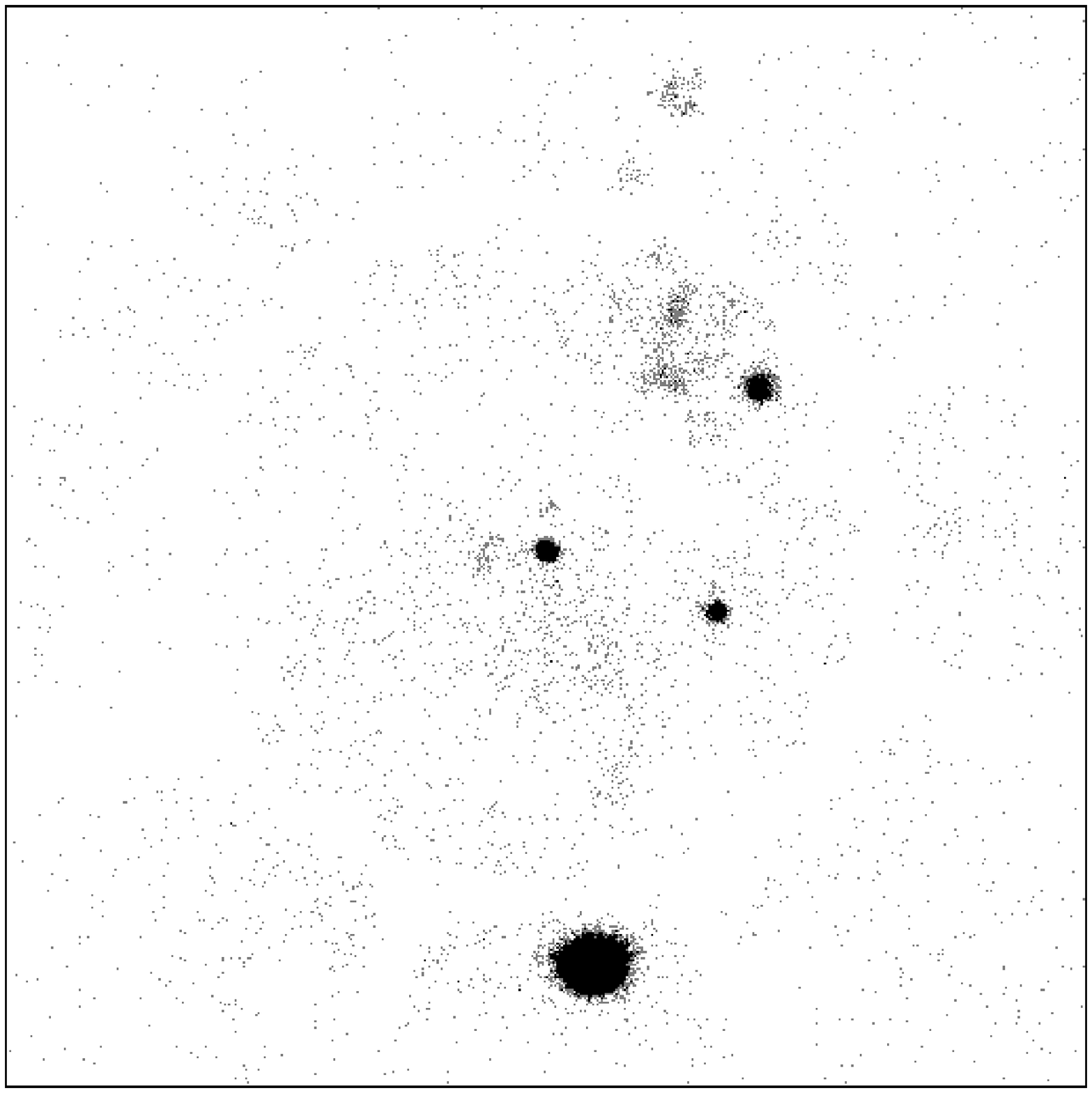}
\includegraphics{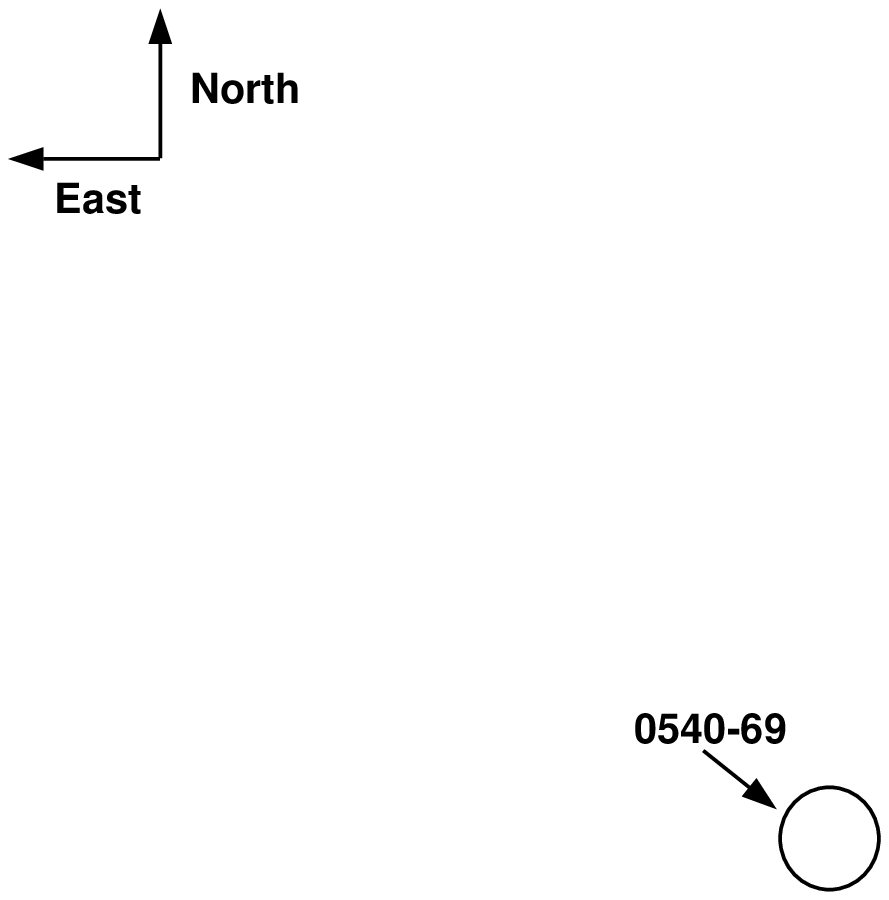}
\caption{Typical ROSAT PSPC image of the Large Magellanic Cloud.  0540-69 is the source at the field center.  LMC X-1 is the bright source at the bottom of the field.}
\end{figure}

\begin{figure}
\vspace*{180mm}
\includegraphics{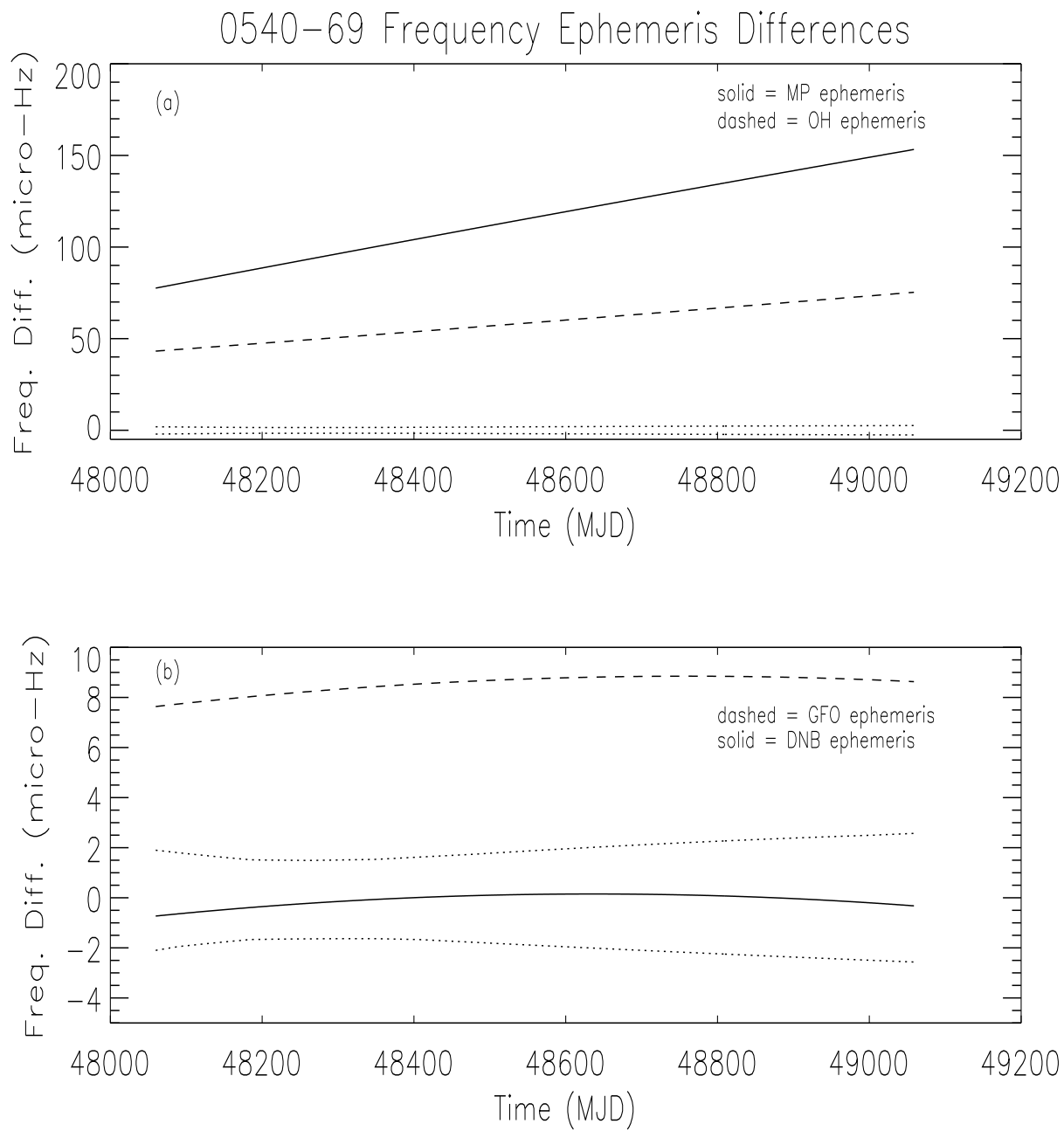}
\caption{Frequency difference between our frequency-based timing solution in Table 4 and other published ephemerides for 0540-69 - (a) Manchester and Peterson (1989) and \"{O}gelman and Hasinger (1990) ephemerides, (b) GF\"{O} and DNB ephemerides.  The dotted lines bound the $\pm 1 \sigma$ range in allowed frequency for our solution.}
\end{figure}

\begin{figure}
\vspace*{130mm}
\includegraphics{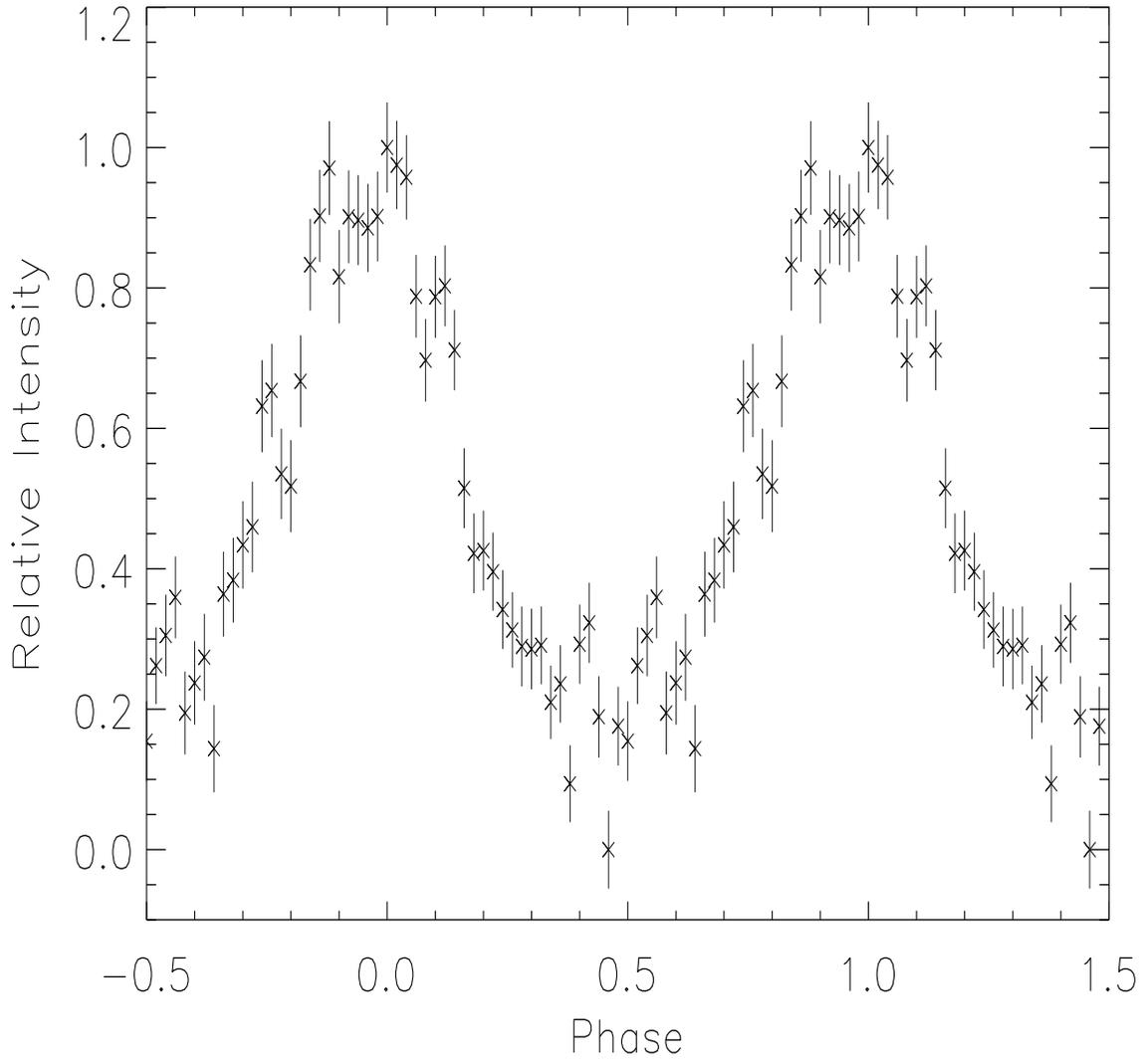}
\caption{ROSAT PSPC master pulse profile for 0540-69}
\end{figure}

\begin{figure}
\vspace*{180mm}
\includegraphics{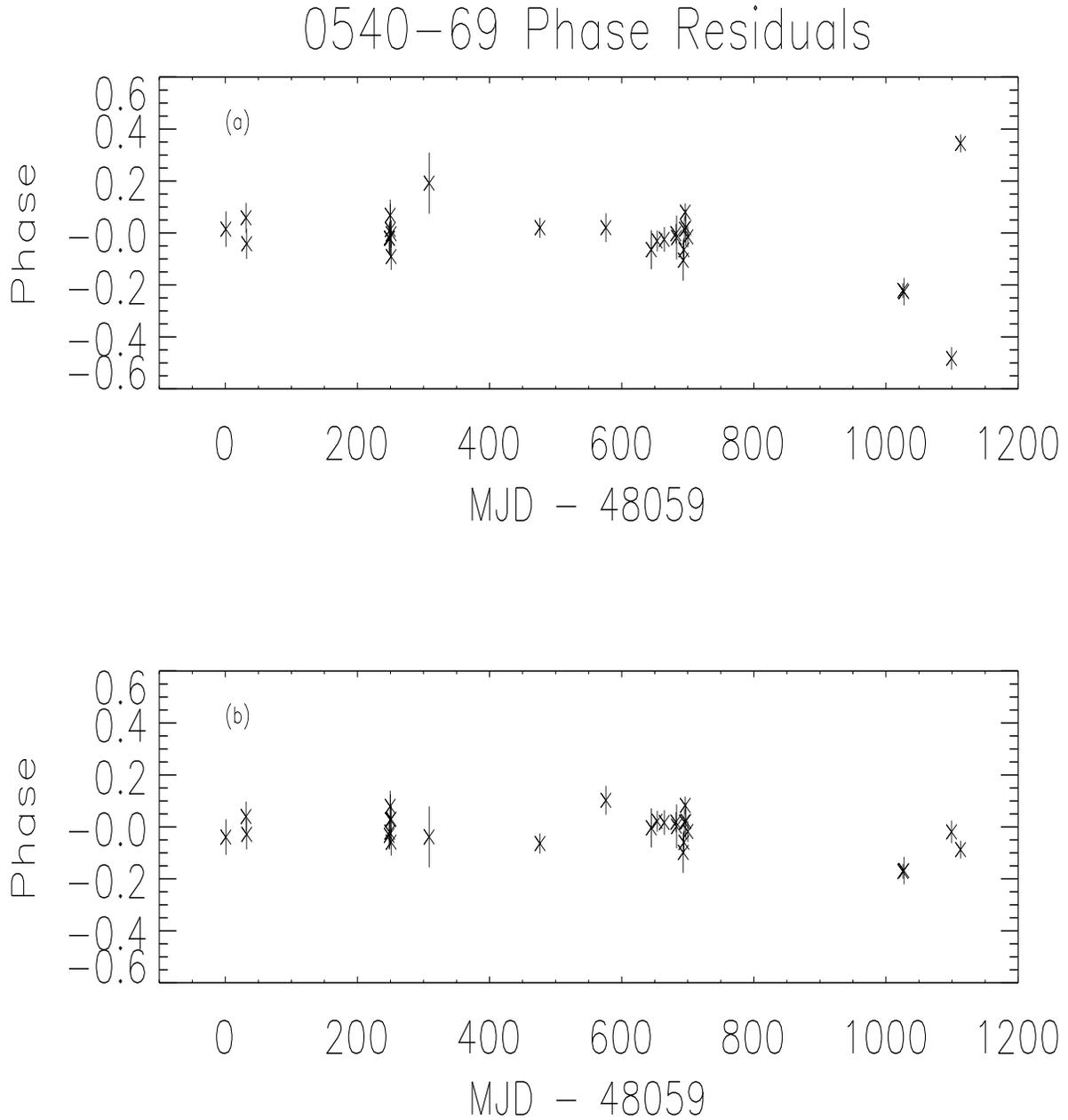}
\caption{Phase residuals for (a) Solution A in Table 5, and (b) Solution B in Table 5.  While both solutions give $\chi^2 \sim 19$ for TOAs 1-23, they have $\chi^2 = 321$ and 74, respectively, for all 27 TOAs.}
\end{figure}

\end{document}